# Scalable Anomaly Ranking of Attributed Neighborhoods

Bryan Perozzi     Leman Akoglu
Stony Brook University
Department of Computer Science
{bperozzi, leman}@cs.stonybrook.edu

**Abstract**

Given a graph with node attributes, what neighborhoods[1] are anomalous? To answer this question, one needs a quality score that utilizes both structure and attributes. Popular existing measures either quantify the structure only and ignore the attributes (e.g., conductance), or only consider the connectedness of the nodes inside the neighborhood and ignore the cross-edges at the boundary (e.g., density).

In this work we propose normality, a new quality measure for attributed neighborhoods. Normality utilizes structure and attributes *together* to quantify both internal consistency and external separability. It exhibits two key advantages over other measures: (1) It allows many boundary-edges as long as they can be "exonerated"; i.e., either (*i*) are expected under a null model, and/or (*ii*) the boundary nodes do not exhibit the subset of attributes shared by the neighborhood members. Existing measures, in contrast, penalize boundary edges irrespectively. (2) Normality can be efficiently maximized to automatically infer the shared attribute subspace (and respective weights) that characterize a neighborhood. This efficient optimization allows us to process graphs with millions of attributes.

We capitalize on our measure to present a novel approach for Anomaly Mining of Entity Neighborhoods (AMEN). Experiments on real-world attributed graphs illustrate the effectiveness of our measure at anomaly detection, outperforming popular approaches including conductance, density, OddBall, and SODA. In addition to anomaly detection, our qualitative analysis demonstrates the utility of normality as a powerful tool to contrast the correlation between structure and attributes across different graphs.

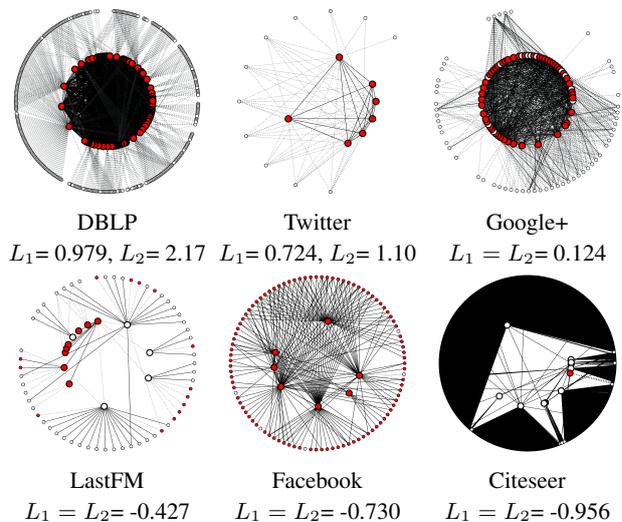

Figure 1: Example neighborhoods (inner circles) and their boundaries (outer circles) in our real-world graphs from high (top-left) to low (bottom-right) normality scores. Colors depict presence (red) or absence (white) of the attribute that maximizes normality for each neighborhood. Dashed edges are "exonerated". Results discussed further in Section 5.1.

## 1 Introduction

Graph anomaly detection [2] is a problem of pressing concern, with broad applications including network security, spam/fraud detection (in social networks, financial networks, etc.), database integrity, and more.

The essence of graph anomaly detection lies in determining an appropriate *anomaly score*, which effectively characterizes the quality of each neighborhood. Quantifying the quality of graph neighborhoods has long been a research area of interest on its own, where it finds additional applications in network community detection and graph partitioning. Most of the existing approaches focus on unattributed or plain graphs and hence only utilize structure.

These structural approaches can be divided by whether they focus on the internal characterizations of a neighborhood, its separability from its boundary, or a combination of both. Measures which solely use internal information judge quality on intra-group connectivity statistics (e.g. average degree, edge density) and have been used for dense subgraph mining [5, 17] and anomaly detection [1]. In contrast to internal measures, boundary-centric measures (e.g. expansion, cut-ratio) define a neighborhood's quality only in terms of how separable it is from the rest of the graph. Perhaps the most popular measures are those which characterize a neighborhood by both its internal characteristics and its boundary (e.g. modularity [15] and conductance [4]).

---
[1] A neighborhood is used as a general term throughout text and refers to any connected subgraph; such as a cluster, community, social circle, ego network, etc.

In contrast to numerous measures that focus solely on structural quality, there exist only a few attempts that also utilize attributes. For example, several methods aim to find communities in attributed graphs that not only are dense but also exhibit attribute coherence [3, 8, 10]. Most recently, others also quantify the connectivity at the boundary to find outlier nodes or subgraphs in attributed graphs [11, 16].

In this work, we propose a novel approach to detecting anomalous neighborhoods in attributed graphs. We define a neighborhood to be high quality when its nodes are (1) internally well connected and similar to each other on a specific attribute subspace (we call these shared attributes the *neighborhood focus*), as well as (2) externally well separated from and/or dissimilar to the nodes at the boundary. Based on this definition, we introduce a new measure called `normality` to quantify the quality of attributed neighborhoods, which carefully utilizes both structure and attributes together to quantify their internal consistency within, as well as external separability at the boundary. (See Figure 1 for examples of high-to-low `normality` neighborhoods from various real-world graphs.) We note that the focus attributes of the neighborhoods may be latent and unknown a priori, especially in high dimensions. Our method AMEN (for Anomaly Mining of Entity Neighborhoods) automatically *infers* the focus attributes and their respective weights, so as to maximize the normality score of a neighborhood. Neighborhoods with low normality scores, i.e., for which a focus that yields high `normality` cannot be found, are considered low quality or anomalous. The main contributions of our work are summarized as follows:

- **Neighborhood Quality Score**: We propose `normality`, a new measure to quantify the quality of the structure (topology) as well as the focus (attributes) of neighborhoods in attributed graphs. Intuitively, `normality` quantifies the extent which a neighborhood is "coherent"; i.e., ($i$) internally consistent and ($ii$) externally separated from its boundary.
- **Neighborhood Anomaly Mining**: We formulate and solve a novel anomaly mining task for attributed graphs. Our proposed method, AMEN, discovers a given neighborhood's *latent* focus through the unsupervised maximization of its `normality`. Those neighborhoods for which a proper focus cannot be identified receive low score, and are deemed as anomalous.
- **Scalable Optimization**: Our proposed measure lends itself to an efficient convex optimization procedure. This allows us to analyze real-world graphs with millions of attributes.

Experiments on real-world attributed graphs show the effectiveness of our approach. Specifically, we show the utility of our measure in spotting anomalies in attributed graphs, where AMEN outperforms existing approaches including conductance, density, OddBall [1], and SODA [11] by 16%-25% mean precision. Furthermore, we qualitatively analyze the high and low `normality` neighborhoods, and show how our method can effectively contrast differences in correlation between structure and attributes across graphs.

## 2 Problem Statement

We consider the ranking problem of entity neighborhoods in a graph with node attributes by quality. More formally, let $G = (\mathcal{V}, \mathcal{E}, \mathcal{A})$ denote an attributed graph with $|\mathcal{V}| = n$ nodes, $|\mathcal{E}| = m$ edges, and $|\mathcal{A}| = d$ node attributes (or features). A neighborhood of $G$ is defined as a set of nodes $C \subseteq \mathcal{V}$ and the edges among them, $(i,j) \in \mathcal{E}, \{i,j\} \in C$. We denote by $B \subseteq \mathcal{V}$ the set of boundary nodes that are outside the neighborhood but have at least one edge to some node in the neighborhood, i.e., $(c,b) \in \mathcal{E}, c \in C, b \in B, C \cap B = \emptyset$.

We consider a neighborhood to be of high quality based on two criteria: ($i$) internal consistency, and ($ii$) external separability. Intuitively, a good neighborhood has many internal edges among its members where they share a set of attributes with similar values. In other words, a common set of *focus* attributes makes the neighborhood members highly similar. In addition, a good neighborhood has either only a few edges at its boundary or many of the cross-edges can be "exonerated", that is, the focus attributes that make the neighborhood members similar to one another also make them dissimilar to or separated from their boundary nodes.

In summary, the *Anomaly Mining of Entity Neighborhoods* (AMEN) problem is given as follows:

**Given** a set of neighborhoods $\mathcal{C} = \{C_1, \ldots, C_k\}$ from a graph $G = (\mathcal{V}, \mathcal{E}, \mathcal{A})$ with node attributes;
**Define** a measure to quantify the quality of $C_i$'s based on internal connectivity, boundary $B_i$, and attributes $\mathcal{A}$,
**Find** the neighborhoods of lowest quality.

## 3 Neighborhood Quality in Attributed Graphs

In this work we propose `normality`, a new measure of neighborhood quality based on a combination of internal connectivity, external separability, and attributes. Our measure is motivated by two properties of real world networks: ($i$) that there are no good cuts in real world graphs [12], and ($ii$) that most user-defined social circles overlap with each other [13]. Together, these imply the existence of (many) cross-edges between neighborhoods and at their boundary. Figure 2 illustrates two specific scenarios that could drive the emergence of many cross-edges at the boundary of high-quality neighborhoods in real-world attributed graphs.

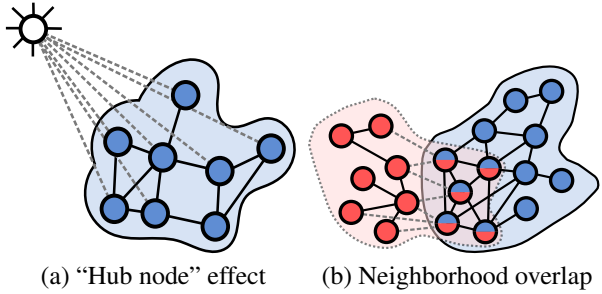

(a) "Hub node" effect  (b) Neighborhood overlap

**Figure 2: High-quality neighborhoods *can* have bad cuts (i.e., many boundary-edges) due to (a) hub nodes or (b) neighborhood overlap. Our `normality` measure carefully utilizes a null model in (a) and node attributes in (b) to exonerate such edges (dashed lines).**

The first scenario is where the cross-edges are due to hub nodes in real graphs. Consider Figure 2 (a) as an example, which shows a well structured neighborhood and a hub node in the host graph. Notice that the hub node connects to a considerable fraction of the nodes in the neighborhood, creating many cross-edges at its boundary. These edges, however, are *not surprising*. In fact, they are expected—as hub nodes by definition connect to a large body of nodes in a graph. While the quality of such a neighborhood is diminished e.g., based on conductance, our measure exonerates those edges as unsurprising under a null model, and does not penalize the neighborhood's `normality` score.

Another scenario where good neighborhoods have many cross-edges at their boundary is when the neighborhoods overlap. An example is given in Figure 2 (b) where two overlapping neighborhoods are shown. Good neighborhoods have many internal edges among their nodes, which implies that overlap nodes have many edges to non-overlap nodes in both neighborhoods. This in turn creates many cross-edges for both neighborhoods at their boundary. These edges, however, are driven by the internal density and should not affect their (external) quality. Provided that overlapping neighborhoods have sufficiently different *focus* that makes them separable (e.g., football vs. chess group), `normality` exonerates such cross-edges based on the attribute dissimilarity of boundary nodes to internal nodes. In contrast, measures that ignore attributes (e.g., cut-ratio, conductance, etc.) penalize these cross-edges irrespectively. Those measures are expected to perform poorly for graphs with many overlapping communities such as social networks.

These scenarios illustrate that using both structure and attributes is meaningful and necessary to quantify the quality of neighborhoods in real graphs. Our `normality` measure is unique in its notion of "exoneration" of edges at the boundary, which none of the existing measures exhibit. Rather, they either completely ignore the boundary or equally penalize all the boundary edges irrespective of context.

**3.1 Proposed Measure** Next we formally introduce `normality`, our new measure to quantify the quality of attributed neighborhoods. `Normality` is inspired by Newman's modularity and assortativity [15, 14] (See Appendix A) but exhibits key differences. First, `normality` utilizes *both structure and attributes* together. Second, its formulation generalizes to graphs with *multiple* node attributes (as opposed to assortativity which is defined only for a single node attribute). Our measure is built upon two intuitive criteria that define a high quality attributed neighborhood, (1) internal consistency and (2) external separability.

**3.1.1 Internal consistency:** To quantify the internal consistency of a neighborhood, we propose to generalize scalar assortativity (Eq. (A.3)) which is defined over a graph with a single attribute to a set of nodes with multiple attributes. The internal consistency $I$ of a neighborhood $C$ is then written as

$$
\begin{aligned}
I = cov(\mathbf{x_i}, \mathbf{x_j}) &= \sum_{f=1}^{|\mathcal{A}|}\bigg(\sum_{i\in C, j\in C}\Big(A_{ij} - \frac{k_i k_j}{2m}\Big)\mathbf{x_i}(f)\mathbf{x_j}(f)\bigg) \\
&= \sum_{i\in C, j\in C}\Big(A_{ij} - \frac{k_i k_j}{2m}\Big)\sum_{f=1}^{|\mathcal{A}|}\mathbf{x_i}(f)\mathbf{x_j}(f) \\
(3.1) \quad &= \sum_{i\in C, j\in C}\Big(A_{ij} - \frac{k_i k_j}{2m}\Big)\ \mathbf{x_i}\cdot\mathbf{x_j}
\end{aligned}
$$

where $A$ is the adjacency matrix of graph $G$, $k_i$ denotes node $i$'s degree, and $\mathbf{x_i}$ is the *vector* of attributes that node $i$ is associated with.

We remark that $\mathbf{x_i}\cdot\mathbf{x_j}$ translates to the dot-product similarity between the attribute vectors of neighborhood members. However, it treats all of the attributes as equally important in quantifying the similarity among nodes. In general, it is more reasonable to assume that the neighborhood members come together around *a few* common attributes (e.g., same school and same hobby). This is expected especially in very high dimensions. We refer to those attributes upon which neighborhood members agree, i.e. have similar values, as the *focus* (attributes) of a neighborhood.

Therefore, we modify the internal consistency score by introducing a non-negative weight vector $\mathbf{w}$ to compute the weighted node similarities as

$$
(3.2) \quad I = \sum_{i\in C, j\in C}\Big(A_{ij} - \frac{k_i k_j}{2m}\Big)\ s(\mathbf{x_i}, \mathbf{x_j}|\mathbf{w})\ ,
$$

for similarity $s(\mathbf{x_i}, \mathbf{x_j}|\mathbf{w}) = \mathbf{w}^T\cdot(\mathbf{x_i}\circ\mathbf{x_j})$, where $\circ$ denotes the Hadamard (element-wise) product, and $\mathbf{w}$ is a vector with the attribute weights. This corresponds to weighted dot-product similarity, $(\mathbf{w}^{1/2}\circ\mathbf{x_i})^T\cdot(\mathbf{w}^{1/2}\circ\mathbf{x_j})$. We expect $\mathbf{w}$ to be *sparse*, in which only a few attributes corresponding to the neighborhood *focus* have large values and zero elsewhere.

Note that each neighborhood $C$ is associated with its own weight vector $\mathbf{w_C}$, i.e. focus, which is potentially different across neighborhoods. Moreover, the attribute weights are often latent. For defining our quality criteria we can assume **w** is known. Later in Section 4 we will show that thanks to our formulation, we can infer this weight vector so as to make a given neighborhood as internally consistent and externally well-separated as possible. In the following we discuss the properties captured by Eq. (3.2).

First, notice that the internal consistency is decreased by missing edges inside a neighborhood, as $A_{ij} = 0$ for $(i,j) \notin \mathcal{E}$. Second, the existence of an edge is rewarded as much as the "surprise" of the edge. Specifically, $\frac{k_i k_j}{2m}$ denotes the probability that two nodes of degrees $k_i$ and $k_j$ are connected to each other by chance in a random network with the same degree distribution as the original graph [15]. As such, we define the surprise of an edge $(i,j) \in \mathcal{E}$ as $(1 - \frac{k_i k_j}{2m})$. The smaller $\frac{k_i k_j}{2m}$ is for an existing edge inside a neighborhood, the more surprising it is and the more it contributes to the quality of the neighborhood.

These two properties quantify the *structure* of the neighborhood. On the other hand, the similarity function quantifies the *attribute* coherence. As a result, the more similar the neighborhood nodes can be made by some choice of **w**, the higher $I$ becomes. If no such weights can be found, internal consistency reduces even if the community is a complete graph with no missing edges.

Overall, a neighborhood with (1) many existing and (2) "surprising" internal edges among its members where (3) (a subset of) attributes make them highly similar receives a high internal consistency score.

**3.1.2 External separability:** Besides being internally consistent, we consider a neighborhood to be of high quality if it is also well-separated from its boundary. In particular, a well-separated neighborhood either has (1) few cross-edges at its boundary, or (2) many cross-edges that can be "exonerated". A cross-edge $(i,b) \in \mathcal{E}$ ($i \in C, b \in B$) is exonerated either when it is unsurprising (i.e., expected under the null model) or when internal node $i$ is dissimilar to boundary node $b$ based on the focus attribute weights. The latter criterion ensures that what makes the neighborhood members similar to one another does not also make them similar to the boundary nodes, but rather differentiates them. The external separability $E$ of a neighborhood $C$ is then

(3.3)
$$E = - \sum_{\substack{i \in C, b \in B, \\ (i,b) \in \mathcal{E}}} \left(1 - \min(1, \frac{k_i k_b}{2m})\right) s(\mathbf{x_i}, \mathbf{x_b} | \mathbf{w}) \leq 0 \;.$$

External separability considers only the boundary edges and quantifies the degree that these cross-edges can be exonerated. As discussed earlier, cross-edges are exonerated in two possible ways. First, a cross-edge may be unsurprising; in which case the term $(1 - \min(1, \frac{k_i k_b}{2m}))$ becomes small or ideally zero (recall Fig. 2 (a) scenario). Second, the boundary node of a cross-edge may not share the same focus attributes with the internal node; in which case the term $s(\mathbf{x_i}, \mathbf{x_b}|\mathbf{w})$ becomes small or ideally zero (recall Fig. 2 (b) scenario). The higher the number of cross-edges that can be exonerated, the larger $E$ (note the negative sign) and hence the quality of a neighborhood becomes.

Note that good neighborhoods by `normality` differ from quasi-cliques for which only internal quality measures, such as density [17] or average degree [5], are defined. Different from those and besides internal consistency, we also quantify the quality of the boundary of a neighborhood. `Normality` is also different from popular measures that do quantify the boundary, such as cut-ratio [7], modularity [15] or conductance [4], for which good neighborhoods are expected to have only a few cross-edges. In contrast, our formulation allows for *many* cross-edges *as long as* they are either $(i)$ unsurprising (under the null model) or if surprising, $(ii)$ can be exonerated by the neighborhood *focus*. These advantages arise as we utilize both structure and attributes in a systematic and intuitive way to define our measure.

**3.2 Normality** Having defined the two criteria for the quality of a neighborhood, `normality` ($N$) is written as the sum of the two quantities $I$ and $E$, where high quality neighborhoods are expected to have both high internal consistency and high external separability.

$$N = I + E = \sum_{i \in C, j \in C} \left(A_{ij} - \frac{k_i k_j}{2m}\right) s(\mathbf{x_i}, \mathbf{x_j}|\mathbf{w})$$

(3.4)
$$- \sum_{\substack{i \in C, b \in B \\ (i,b) \in \mathcal{E}}} \left(1 - \min(1, \frac{k_i k_b}{2m})\right) s(\mathbf{x_i}, \mathbf{x_b}|\mathbf{w})$$

For a neighborhood with the highest `normality`, all the possible internal edges exist and are also surprising for which pairwise similarities are high. These ensure that the first term is maximized. Moreover, the neighborhood either has no cross-edges or the similarity or surprise of existing cross-edges to the boundary nodes are near zero, such that the second term vanishes. Neighborhoods of a graph for which the `normality` takes negative values are of lesser quality and deemed as anomalous.

*Choice of similarity function:* To this end, we considered the node attributes to be scalar variables where $s(\mathbf{x_i}, \mathbf{x_j}|\mathbf{w})$ is the weighted dot-product similarity. If the attributes are categorical (e.g., location, occupation, etc.), one can instead use the Kronecker delta function $\delta(\cdot)$ that takes the value 1 if two nodes exhibit the same value for a categorical attribute and 0 otherwise.

The choice of the similarity function is especially important for binary attributes (e.g., likes-biking, has-job, etc.). While those can be thought of as categorical variables taking the values $\{0,1\}$, using Kronecker $\delta$ becomes undesirable for nodes inside a neighborhood. The reason is, inter-nal consistency by the $\delta$ function is the same both when all the neighborhood nodes exhibit a particular binary attribute (all 1) and when none does (all 0). However, one may not want to characterize a neighborhood based on attributes that its members do not exhibit even if the agreement is large. Therefore, we propose to use dot-product for computing internal consistency and Kronecker $\delta$ for computing external separability for binary-attributed graphs.

## 4 Anomaly Mining of Entity Neighborhoods

As presented so far, when given a neighborhood $C$ of an attributed graph and vector $\mathbf{w}$ of attribute weights, we can directly compute its `normality` using Eq. (3.4). However, for the task of anomaly mining, the *focus* of a neighborhood is latent and hard to guess without any prior knowledge. This is especially true in high dimensions where most attributes are irrelevant, making a uniform attribute weight vector impractical. Moreover, even if the neighborhood focus is known a priori, it is hard to assign weights to those attributes beyond that of binary relevance.

In this section, we propose an optimization approach to automatically *infer* the attribute weight vector for a given neighborhood, as the vector that maximizes its normality score. That is, we aim to identify a subspace that would make the neighborhood's `normality` as high as possible. All neighborhoods can then be ranked based on their (best possible) normality scores, and those with lowest scores can be deemed anomalous. This allows us to restate our original problem in Section 2 as follows:

**Given** a set of neighborhoods $\mathcal{C}$ and `normality` $N$;
**Find** the attribute weight vector $\mathbf{w}_{\mathbf{C_i}}$ which maximizes $N(C_i)$ for each neighborhood $C_i \in \mathcal{C}$,
**Rank** neighborhoods $\mathcal{C}$ by normality score,
**Find** the neighborhoods of lowest quality.

**4.1 Neighborhood Focus Extraction** Our goal is to find an attribute weight vector (hereafter called $\mathbf{w_C}$) for a neighborhood $C$, which makes its `normality` as high as possible, such that connected nodes in the neighborhood are very similar and the nodes at the boundary are dissimilar. To this end, we leverage our `normality` to formulate an objective function parameterized by the attribute weights. This objective also has the nice property of quantifying structure, by penalizing non-existing in-edges and surprising cross-edges. Our formulation for focus extraction is then $\max_{\mathbf{w_C}} N(C)$, which by reorganizing the terms that do not depend on $\mathbf{w_C}$, can be rewritten (based on Eq. (3.4)) as

$$\max_{\mathbf{w_C}} \mathbf{w_C}^T \cdot \Big[ \sum_{i \in C, j \in C} \big(A_{ij} - \frac{k_i k_j}{2m}\big) s(\mathbf{x_i}, \mathbf{x_j}) - \sum_{\substack{i \in C, b \in B \\ (i,b) \in \mathcal{E}}} \big(1 - \min(1, \frac{k_i k_b}{2m})\big) s(\mathbf{x_i}, \mathbf{x_b}) \Big]$$

(4.5) $$\max_{\mathbf{w_C}} \mathbf{w_C}^T \cdot (\mathbf{x}_I + \mathbf{x}_E)$$

where $\mathbf{x}_I$ and $\mathbf{x}_E$ are vectors that respectively denote the first and the second summation terms. Note that these can be directly computed from data. Moreover, the similarity function $s(\mathbf{x_i}, \mathbf{x_j})$ can be replaced by either $(\mathbf{x_i} \circ \mathbf{x_j})$ or $\delta(\mathbf{x_i}, \mathbf{x_j})$ depending on the type of the node attributes.

**4.2 Size-invariant Scoring** The normality score in Eq. (4.5) grows in magnitude with the size of the neighborhood $C$ being considered. Normalization is desirable then, in order to compare across differences in both neighborhood and boundary size.

We note that $I$ is the maximum in the case of a fully connected neighborhood the members of which all agree upon the focus attributes. Therefore, $I_{\max} = |C|^2$, where $s_{\max}(\mathbf{x_i}, \mathbf{x_j}) = 1$ provided that the attributes $\mathbf{x_i}(f)$ are normalized to $[0, 1]$ for each node $i$. On the other hand the minimum is negative, when there exists no internal edges and pairwise similarities are maximum. That is, $I_{\min} = \sum_{i \in C, j \in C} -\frac{k_i k_j}{2m}$. To normalize the internal consistency $I$, we subtract $I_{\min}$ and divide by $I_{\max} - I_{\min}$, which is equivalent to a weighted version of edge density.

To normalize external separability, we derive a measure similar to conductance [4], i.e., ratio of boundary or cut edges to the total volume (sum of the degrees of the neighborhood nodes). The difference is that each edge is weighted based on its surprise and the similarity of its end nodes. In particular, we define $\mathbf{x}_{\tilde{I}} = \sum_{\substack{i \in C, j \in C \\ (i,j) \in \mathcal{E}}} \big(1 - \min(1, \frac{k_i k_j}{2m})\big) s(\mathbf{x_i}, \mathbf{x_j})$. Note that similar to $E$, $\tilde{I}$ considers only the existing edges in the graph. Therefore, $\tilde{I} - E$ can be seen as the total weighted volume of the neighborhood.

Overall, we scale our measure as follows, where the division of the vectors in the second term is element-wise. As such, $\hat{\mathbf{x}}_I(f) \in [0, 1]$ and $\hat{\mathbf{x}}_E(f) \in [-1, 0]$.

$$\hat{N} = \mathbf{w_C}^T(\hat{\mathbf{x}}_I + \hat{\mathbf{x}}_E) = \mathbf{w_C}^T \big(\frac{\mathbf{x}_I - I_{\min}}{I_{\max} - I_{\min}} + \frac{\mathbf{x}_E}{\mathbf{x}_{\tilde{I}} - \mathbf{x}_E}\big)$$

**4.3 Objective Optimization** The normalized objective function can be written as
(4.6)
$$\max_{\mathbf{w_C}} \mathbf{w_C}^T \cdot (\hat{\mathbf{x}}_I + \hat{\mathbf{x}}_E)$$
$$\text{s.t.} \quad \|\mathbf{w_C}\|_p = 1, \ \mathbf{w_C}(f) \geq 0, \ \forall f = 1 \ldots d$$

Note that we introduce a set of constraints on $\mathbf{w_C}$ to fully formulate the objective. In particular, we require the attribute weights to be non-negative and that $\mathbf{w_C}$ is normalized (or regularized) to its $p$-norm. These constraints also facilitate the interpretation of the weights. In the following we let $\mathbf{x} = (\hat{\mathbf{x}}_I + \hat{\mathbf{x}}_E)$, where $\mathbf{x}(f) \in [-1, 1]$.

There are various ways to choose $p$, yielding different interpretations. If one uses $\|\mathbf{w_C}\|_{p=1}$, a.k.a. the $L_1$ norm,

Table 1: Real-world graphs used in this work. * depicts datasets with ground truth circles. $n$: number of nodes, $m$: number of edges, $d$: number of attributes, $|\mathcal{C}|$: number of circles, $|S|$: average circle size.

| Name | $n = |\mathcal{V}|$ | $m = |\mathcal{E}|$ | $d = |\mathcal{A}|$ | $|\mathcal{C}|$ | $|S|$ | nodes | edges | attributes |
|---|---|---|---|---|---|---|---|---|
| *Facebook | 4,039 | 88,234 | 42-576 | 193 | 21.93 | users | friendships | user profile information |
| *Twitter | 81,306 | 1,768,149 | 1-2,271 | 4,869 | 12.51 | users | follow relations | hashtags and user mentions |
| *Google+ | 107,614 | 13,673,453 | 1-4,122 | 479 | 134.75 | users | friendships | user profile information |
| DBLP | 108,030 | 276,658 | 23,285 | n/a | n/a | authors | co-authorships | title words used in articles |
| Citeseer | 294,104 | 782,147 | 206,430 | n/a | n/a | articles | citations | abstract words used in articles |
| LastFM | 272,412 | 350,239 | 3,929,101 | n/a | n/a | users | friendships | music pieces listened to |

the solution picks as the neighborhood focus the *single* attribute with the largest $\mathbf{x}$ entry. That is, $\mathbf{w_C}(f) = 1$ where $\max(\mathbf{x}) = \mathbf{x}(f)$ and 0 otherwise. One can interpret this as the most important attribute that characterizes the neighborhood. Note that $\mathbf{x}$ may contain only negative entries, in which case the largest negative entry is selected. This implies that there exists no attribute that can make the `normality` positive, and hence such a neighborhood is considered anomalous. Note that when $p = 1$, $\hat{N} \in [-1, 1]$.

If there are *multiple* attributes with positive $\mathbf{x}$ entries, we can also select all of them as the neighborhood focus. The weights of these attributes, however, should be proportional to the magnitude of their $\mathbf{x}$ values. This is exactly what $\|\mathbf{w_C}\|_{p=2}$, or the $L_2$ norm yields. In particular, we can show that $\mathbf{w_C}(f) = \frac{\mathbf{x}(f)}{\sqrt{\sum_{\mathbf{x}(i)>0} \mathbf{x}(i)^2}}$, for $\mathbf{x}(f) > 0$ and 0 otherwise, where $\mathbf{w_C}$ is unit-normalized. Then, the normality score of the neighborhood becomes

$$N = \mathbf{w_C}^T \cdot \mathbf{x}$$
$$= \sum_{\mathbf{x}(f)>0} \frac{\mathbf{x}(f)}{\sqrt{\sum_{\mathbf{x}(i)>0} \mathbf{x}(i)^2}} \mathbf{x}(f) = \sqrt{\sum_{\mathbf{x}(i)>0} \mathbf{x}(i)^2} = \|\mathbf{x}_+\|_2$$

i.e., the $L_2$-norm of $\mathbf{x}$ induced on the positive entries. As such, when there are multiple attributes that can make the `normality` positive, $L_2$ formulation produces an objective value that is higher than that of the $L_1$ formulation. This agrees with intuition; the larger the number of attributes with positive $\mathbf{x}$ entries, the more attribute-coherence the neighborhood exhibits, and the higher the `normality` gets incrementally. On the other hand, if there are no positive entries in $\mathbf{x}$, the $L_2$ optimization selects the single attribute with the largest negative entry, and we consider the neighborhood as anomalous. In all, $\hat{N} \in [-1, \|\mathbf{x}_+\|_2]$ when $p = 2$.

While $L_1$ and $L_2$ are the two most commonly used norms, one can also enforce $\mathbf{w_C}(f) \leq \frac{1}{k}$, for each $f$, to obtain the largest $k$ entries of $\mathbf{x}$ that can be interpreted as the top-$k$ most relevant attributes for the neighborhood (note that those may involve both positive and negative entries). In principle, $\mathbf{x}$ provides a systematic and intuitive way to rank the attributes by their relevance to a neighborhood.

*Computational complexity:* Notice that the solution to the optimization is quite straightforward where the complexity mainly revolves around computing the $\mathbf{x}$ vector. Specifically, the complexity is $O(|C|^2 d + |\mathcal{E}_B|d)$ for computing $\mathbf{x}$ and $O(d)$ for finding the maximum entry (for $L_1$ regularization) or positive entries (for $L_2$ regularization), where $\mathcal{E}_B$ is the number of cross-edges which is upper-bounded by $|C||B|$. Therefore, the complexity is quadratic w.r.t. the neighborhood size $|C| \ll n$, and *linear* in the number of attributes $d$, while it is independent of the size of the entire graph.

## 5 Experiments

Through experiments, we (1) evaluate AMEN's performance [2] in anomaly detection, (2) perform case studies that analyze the type of anomalies we find, and (3) utilize `normality` as an exploratory tool to study the correlation between structure and attributes across different graphs.

**Datasets.** A detailed description of real-world graphs used in this work is given in Table 1.[3] Facebook, Twitter, and Google+ each consists of a collection of ground-truth social circles. For these graphs, we consider these circles as their entity neighborhoods. For the other graphs, we consider the egonets (subgraphs induced on each node and its neighbors) as their entity neighborhoods.

**Baselines.** We compare AMEN's performance in anomaly detection against existing measures and methods: *Average-Degree* [5], *Cut-Ratio* [7], *Conductance* [4], *Flake-ODF* [6], *OddBall* [1], *SODA* [11], and *Attribute-Weighted Normalized Cut* (AW-NCut) [10] (See Appendix B for definitions).

**5.1 Anomaly Detection** Our evaluation of AMEN's anomaly detection performance is two-fold. First, we perform quantitative evaluation; we inject anomalies into DBLP, Citeseer, and LastFM and compare detection performance of different approaches. Second, we perform a qualitative case study of ground-truth neighborhoods with low `normality` score from Facebook, Twitter, and Google+.

**Quantitative Evaluation.** To create ground truth anomalies, we use the egonets from DBLP, Citeseer, and LastFM which we perturb to obtain anomalous neighborhoods. Perturbations involve disruptions in (1) structure, (2) attributes, and (3) both. We start by choosing "good" neighborhoods; specifically, small egonets (of size 30-100) that we expect to have low conductance cuts [9]. From these egonets, we choose 5% of them as anomalous, i.e., to be perturbed. To perturb structure, we rewire inside edges to random outside nodes with rewiring probability $p$. To perturb attributes, we replace the attributes of inside nodes with the corresponding attributes of randomly picked outside nodes with probability $q$ (note that this "inheritance" only affects the egonet, and keeps the outside nodes unchanged). For structure and attribute perturbation, we vary $p$ or $q$ from 0.05 to 0.50. To perturb both, we vary them simultaneously. The larger the perturbation intensities $p$ and $q$, the more disrupted the egonet. We expect this process to create anomalous (ground truth) egonets, that are structurally poor, for which it is hard to find shared focus attributes.

---
[2]AMEN available at http://www.perozzi.net/projects/amen
[3]Datasets available from http://snap.stanford.edu/data/ and https://code.google.com/p/scpm/.

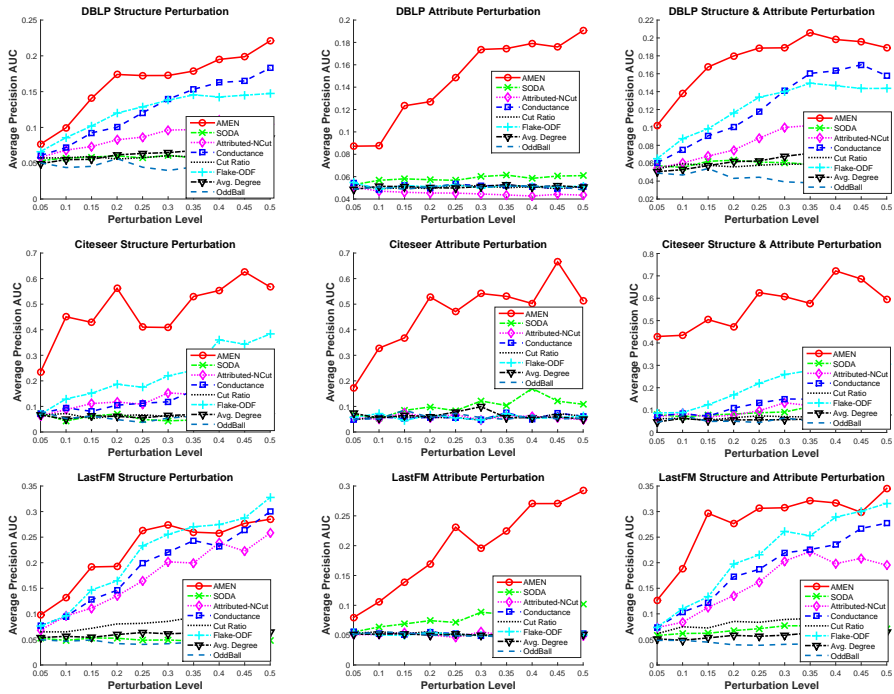

Figure 3: Anomaly detection results (mean precision vs. perturbation intensity) for structure only (left), attribute only (center), and structure & attribute (right) anomalies on DBLP (top), Citeseer (middle), and LastFM (bottom). AMEN is superior, especially when attribute perturbations are involved.

We evaluate our measure in its ability to rank the disrupted ground truth anomalies high. Specifically, we rank the neighborhoods by their normality and report the AUC (i.e., average precision) of the precision-recall plots for each $p$ and/or $q$ perturbation intensity in Figure 3. We find that AMEN consistently outperforms all other measures and methods, especially at low perturbation intensities where the task is harder—hence the gradual increase in performance by intensity. When structure perturbation is involved (See figures on (left) and (right)), conductance and Flake-ODF also appear to do well. When attributes are perturbed, on the other hand, AMEN is superior, where SODA, the second best, is significantly worse than AMEN. Across perturbation strategies and datasets, on average AMEN outperforms Flake-ODF by 16%, conductance by 18%, AW-NCut by 20%, SODA by 23%, average degree by 24%, cut ratio by 24%, and OddBall by 25%.

**Case Studies.** We begin our case studies by returning to Figure 1, which illustrated several examples of neighborhoods sorted by their normality (from high to low) from various graphs. We see that the highest quality neighborhood shown (from DBLP) has good internal structure, and attributes which allow the complete exoneration of edges at its boundary. As we progress through neighborhoods in order of quality, we see that internal structure weakens, and boundaries become larger. The lowest quality neighborhood shown (from Citeseer) has almost no internal structure, and no attributes which exonerate its large boundary.

Next, we examine some of the most anomalous ground truth circles from Facebook, Twitter, and Google+. Figure 4 shows that the circles with the lowest normality have both weak internal connectivity and poor boundary separation. None of the available attributes is able to meaningfully improve their score. We find these ground-truth neighborhoods to be particularly interesting, as they are real circles manually defined by social network users (hence ground truth) which do not, however, exhibit characteristics of what 'good' communities in graphs look like.

Finally, we examine the difference between normality and conductance, a popular structural measure, in Figure 5. As expected, normality gets lower for neighborhoods with increasingly high (bad) conductance (results are for DBLP and Google+, others are similar). However, notice the many neighborhoods in DBLP with high normality scores even if they have very high (in range $(0.9, 1.0]$) conductance (several others in other graphs, and even one in Google+). These are exactly the type of neighborhoods that are considered low quality by solely structural measures such as conductance, but achieve high score when attributes and surprise are carefully accounted for under the notion of "exoneration".

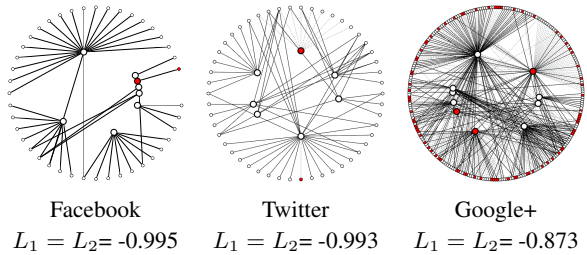

| Facebook | Twitter | Google+ |
|---|---|---|
| $L_1 = L_2 = -0.995$ | $L_1 = L_2 = -0.993$ | $L_1 = L_2 = -0.873$ |

**Figure 4: Low** `normality` **ground truth neighborhoods from Facebook, Twitter, and Google+.**

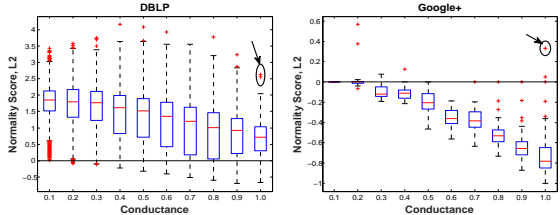

**Figure 5: Box plots depicting** $L_2$-`normality` **score vs. conductance for (a) DBLP and (b) Google+.**

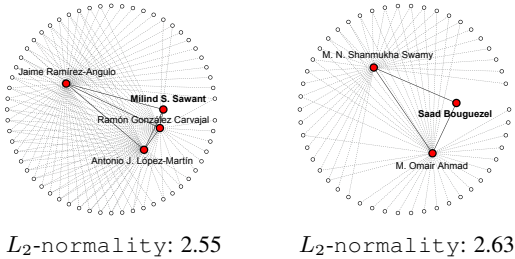

| $L_2$-`normality`: 2.55 | $L_2$-`normality`: 2.63 |
|---|---|
| Focus: {telescopic, multidecade, cascode} | {reciprocal, split, reverse} |

**Figure 6: Example high (poor) conductance but high** `normality` **neighborhoods from DBLP.**

For example, consider the two neighborhoods with poor conductance but high `normality` from DBLP as shown in Figure 6. The left drawing depicts the egonet of 'Milind S. Sawant' (ego)—a co-authorship circle of four researchers from electrical engineering—as well as their boundary (other collaborators). We notice that the other nodes in the egonet, i.e. the neighbors of the ego, have considerably larger set of collaborators. This creates numerous cross-edges, yielding poor conductance. However, this circle is well-defined and "focused"; these four authors exclusively work together on 'telescopic op-amps'; notice the listed focus attributes (words). Similarly, the ego 'Saad Bouguezel' of the right neighborhood is much less connected than his collaborators, creating excessive cross-edges. Those edges are exonerated by `normality`, due to their discriminating focus on 'reciprocal transforms' and 'split-radix FFTs'.

**5.2 Graph Analysis with Normality** `Normality` also serves as a powerful tool for analyzing the correlation of structure and attributes in a graph. In this section, we compare the distribution of neighborhood scores across all graphs. Using `normality` as a lens, we see that:

- Different graphs display distinct distributional fingerprints, indicating very diverse correlation patterns between structure and attributes.
- In all graphs, the biggest gains in `normality` come only from a few attributes—suggesting that a neighborhood *focus* is often sparse.

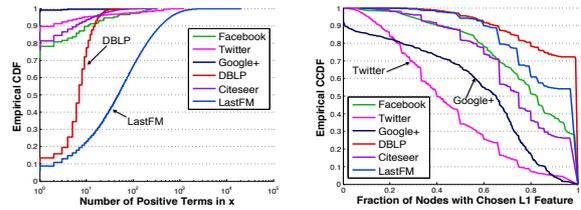

**Figure 7: Distribution of neighborhoods w.r.t. (a) count of positive terms in x (cdf), (b) fraction of nodes exhibiting the** $L_1$-**selected attribute (ccdf).**

First, to examine the influence of attributes upon each neighborhood, we turn to $\mathbf{x} = (\hat{\mathbf{x}}_I + \hat{\mathbf{x}}_E)$ (Eq. (4.6)). For each neighborhood, we count the number of positive entries of $\mathbf{x}$, $\#(\mathbf{x}(i) > 0)$, and present their distribution in Figure 7(a). As expected, only a small number of attributes are relevant for most communities, so a sparse $\mathbf{w}$ is obtained by $L_2$. We see that both LastFM and DBLP have many positive attributes for each neighborhood, followed by Facebook, Citeseer, and Twitter. The worst neighborhoods are found in Google+, where 99% of the circles do not have *any* attribute that characterizes them (i.e., non-positive $\mathbf{x}$). Figure 7(b) shows the fraction of nodes in each neighborhood which exhibit the attribute that would be chosen by $L_1$ maximization. Again, DBLP and LastFM have the most agreement inside each neighborhood, followed closely by Facebook and Citeseer, while Twitter and Google+ neighborhoods are the most noisy in exhibiting the highest ranked attribute.

Figure 8 shows the distribution of normality scores across neighborhoods for both $L_1$ (a) and $L_2$ (b) maximization. We see that 89% of DBLP and 95% of LastFM neighborhoods have positive `normality`, while with the exception of a few very good neighborhoods, the rest are mostly negative. $L_2$ optimization does not dramatically change where the bulk of the distribution is. Majority of DBLP neighborhoods gain additional score, and 10% of LastFM neighborhoods gain a lot of score. In others we see a small improvement, where Google+ presents neighborhoods with lowest `normality` (its best circle score by $L_2$ is $\approx 0.6$).

The contribution from each positive attribute to the neighborhood score for DBLP and LastFM is shown in Figure 9. We see that relevance drops fast, and essentially zeroes out after around 20 attributes. An interesting difference occurs between the two graphs, which have both many good neighborhoods, but achieve them in different ways. In DBLP, the first few positive attributes are all about equally good, but then they degrade quickly. On the other hand, LastFM attributes are usually not as good, but many of them can add up to achieve a neighborhood with a high score (and hence the long tail in Figure 8(b)).

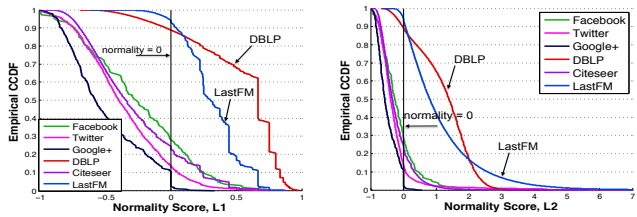

**Figure 8:** Distribution of neighborhoods w.r.t. normality(**ccdf**); w constraint by (a) $L_1$, (b) $L_2$.

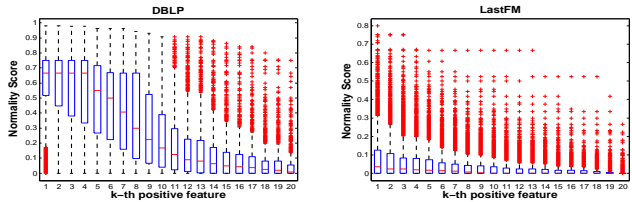

**Figure 9:** Normality of neighborhoods based on $k$-th most positive attribute with highest x entry.

## 6 Conclusion

In this work, we considered the problem of discovering anomalous neighborhoods in attributed graphs. We proposed normality, a new quality measure that evaluates neighborhoods both *I*nternally and *E*xternally. Intuitively, a high-quality neighborhood has members with ($I1$) many *surprising* edges connecting them that ($I2$) share similar values in a particular attribute subspace, called the neighborhood *focus*. Moreover, it has either ($E1$) a few edges at the boundary, or ($E2$) many cross-edges which can be *exonerated* as unsurprising under the null graph model and/or dissimilar with respect to the focus attributes.

We utilized normality within an objective formulation for anomaly mining and provided solutions to maximize it, which automatically identify the latent focus attributes as well as their respective weights. Our formulation yields a scalable optimization that is only quadratic w.r.t. (often small) neighborhood size and linear in attribute size. Overall, normality is unique in its ($i$) exoneration of cross-edges, and ($ii$) automatic inference of focus attributes under a scalable convex optimization.

Experiments on real-world graphs demonstrate the utility of our measure in ranking neighborhoods by quality, where we outperform well-established measures and methods. Normality also provides a powerful tool for studying the correlation between structure and attributes for graphs. Our work enables a number of future directions, including utilizing normality for community detection at large and for user profiling, i.e., recovering missing node attributes.

**Acknowledgments**

The authors thank the anonymous reviewers for their useful comments. This material is based upon work supported by the ARO Young Investigator Program under Contract No. W911NF-14-1-0029, NSF CAREER 1452425, IIS 1408287 and IIP1069147, DARPA Transparent Computing Program under Contract No. FA8650-15-C-7561, a Facebook Faculty Gift, a R&D grant from Northrop Grumman Aerospace Systems, the Institute for Advanced Computational Science at Stony Brook University, and the Stony Brook University Office of Vice President for Research. Any conclusions expressed in this material are of the authors' and do not necessarily reflect the views, either expressed or implied, of the funding parties.

## References

[1] L. Akoglu, M. McGlohon, and C. Faloutsos. Oddball: Spotting anomalies in weighted graphs. In *PAKDD*, 2010.

[2] L. Akoglu, H. Tong, and D. Koutra. Graph-based anomaly detection and description: A survey. *DAMI*, 28(4), 2014.

[3] L. Akoglu, H. Tong, B. Meeder, and C. Faloutsos. PICS: Parameter-free identification of cohesive subgroups in large attributed graphs. In *SIAM SDM*, pages 439–450, 2012.

[4] R. Andersen, F. Chung, and K. Lang. Local graph partitioning using pagerank vectors. In *FOCS*, 2006.

[5] M. Charikar. Greedy approximation algorithms for finding dense components in a graph. In *APPROX*, 2000.

[6] G. W. Flake, S. Lawrence, and C. L. Giles. Efficient identification of web communities. In *KDD*, 2000.

[7] S. Fortunato. Community detection in graphs. *Physics Reports*, 486(3):75–174, 2010.

[8] E. Galbrun, A. Gionis, and N. Tatti. Overlapping community detection in labeled graphs. *DAMI*, 28(5), 2014.

[9] D. F. Gleich and C. Seshadhri. Vertex neighborhoods, low conductance cuts, and good seeds for local community methods. In *KDD*, pages 597–605, 2012.

[10] S. Gunnemann, I. Farber, S. Raubach, and T. Seidl. Spectral subspace clustering for graphs with feature vectors. In *ICDM*, pages 231–240, 2013.

[11] M. Gupta, A. Mallya, S. Roy, J. H. D. Cho, and J. Han. Local learning for mining outlier subgraphs from network datasets. In *SIAM SDM*, pages 73–81, 2014.

[12] J. Leskovec, K. J. Lang, A. Dasgupta, and M. W. Mahoney. Statistical properties of community structure in large social and information networks. In *WWW*, pages 695–704, 2008.

[13] J. J. McAuley and J. Leskovec. Discovering social circles in ego networks. *TKDD*, 8(1):4, 2014.

[14] M. E. J. Newman. Assortative mixing in networks. *Physical Review Letters*, 89(20), 2002.

[15] M. E. J. Newman. Modularity and community structure in networks. *PNAS*, 103(23):8577–8582, 2006.

[16] B. Perozzi, L. Akoglu, P. Iglesias Sánchez, and E. Müller. Focused clustering and outlier detection in large attributed graphs. In *KDD*, pages 1346–1355, 2014.

[17] C. E. Tsourakakis, F. Bonchi, A. Gionis, F. Gullo, and M. A. Tsiarli. Denser than the densest subgraph: extracting optimal quasi-cliques with quality guarantees. In *KDD*, 2013.

# Appendix

## A Preliminaries

Our `normality` measure is inspired by modularity and assortativity, which we describe here.

**A.1 Modularity** Newman's modularity [37] is a measure of network structure that quantifies the extent which the network divides into modules (a.k.a. clusters, communities, groups, circles, etc.). Networks with high modularity have dense connectivity among the nodes within communities, but sparse connectivity between nodes from different communities. Specifically, modularity is written as

$$\text{(A.1)} \qquad M = \frac{1}{2m}\sum_{ij}(A_{ij} - \frac{k_i k_j}{2m})\delta(c_i, c_j) ,$$

where $A$ is the adjacency matrix of graph $G$, $c_i$ denotes the community assignment of node $i$, $k_i$ denotes its degree, and $\delta(\cdot)$ is the indicator function, with value 1 if two nodes belong to the same community and 0 otherwise. Modularity then, is the difference between the actual and the expected fraction of edges between nodes in the same community. The larger the difference, the more modular is the network.

**A.2 Assortativity** While modularity is defined for non-attributed graphs and solely quantifies structure, a similar formula called *assortativity* has been used to measure homophily in attributed networks [36]. Homophily is the extent which the same type of nodes connect to one another, e.g., in social networks [32]. Specifically, for a graph in which every node is associated with a single, *nominal/categorical* attribute (e.g., gender, nationality, etc.), its assortativity is

$$\text{(A.2)} \qquad H^{(\text{nom})} = \frac{1}{2m}\sum_{ij}(A_{ij} - \frac{k_i k_j}{2m})\delta(a_i, a_j) ,$$

where this time $a_i$ depicts the attribute value of node $i$ (e.g., what nationality $i$ belongs to). As such, assortativity is the difference between the actual and the expected fraction of edges between nodes of the same type.

In a perfectly mixed network, all edges fall between nodes of the same type, and assortativity is maximum. It takes negative values for disassortative networks, and is 0 for networks in which attributes and structure are uncorrelated.

**A.3 Scalar Assortativity** Eq. (A.2) is for networks with a nominal/categorical attribute $a$, such as gender, nationality, etc. For a *numerical/scalar* attribute $x$, like income, age, etc., one can derive a corresponding formula using the covariance of the attribute values among connected nodes. Specifically, $cov(x_i, x_j) = \frac{1}{2m}\sum_{ij}A_{ij}(x_i - \mu)(x_j - \mu)$, where $\mu = \frac{1}{2m}\sum_i k_i x_i$ is the mean value of attribute $x$ over the edge-ends, and $k_i$ denotes the degree of node $i$ (note that the average here is over the edges rather than the nodes). From $cov(x_i, x_j)$ one can derive the assortativity for numerically attributed networks (See [38] p.228) as

$$\text{(A.3)} \quad cov(x_i, x_j) = H^{(\text{num})} = \frac{1}{2m}\sum_{ij}(A_{ij} - \frac{k_i k_j}{2m})x_i x_j ,$$

where assortativity is positive when $x_i, x_j$ are both small or both large (w.r.t. the mean), and is negative if they vary in opposite directions. Zero assortativity means the attributes of connected nodes are uncorrelated.

**A.4 Modularity vs. Assortativity** The specific applications that leverage these two measures have traditionally been different. Modularity is often used as an objective function in community detection and graph partitioning [20, 21, 37, 41]. Assortativity, on the other hand, has often been used in measuring homophily in social science studies, e.g., in analyzing how school children of different races and genders interact [33], and how people from various nationalities are segregated in residential areas [30].

Nevertheless, despite the differences between modularity and assortativity, the two quantities are related. It was observed that assortative networks are likely more modular and tend to break into communities in which "like is connected to like" [35]. In other words, one can think of assortativity as a driving force of modular structure in networks—one that influences the emergence of communities.

## B Baselines

We compare our performance in anomaly detection against the following existing measures and methods. Notation used in baselines: $\mathcal{E}(C) = \{(i,j) \in \mathcal{E} : i \in C, j \in C\}$ (edges induced by $C$); $cut(C) = \sum_{i \in C, b \in B, (i,b) \in \mathcal{E}} 1$ (cut size induced by $C$); $vol(C) = \sum_{i \in C} k_i$ (sum of degrees in $C$).

- **Average degree** [5], $\frac{2|\mathcal{E}(C)|}{|C|}$ (internal consistency only, non-attributed).
- **Cut ratio** [7], $\frac{cut(C)}{|C|(n-|C|)}$, is the fraction of boundary edges over all possible boundary edges (external separability only, non-attributed).
- **Conductance** [4], $\frac{cut(C)}{\min(vol(C), vol(G \setminus C))}$, normalizes the cut by the total volume of $C$ (internal+external quality, non-attributed).
- **Flake-ODF** [6], $\frac{\left|\{i \in C : |\{(i,j) \in \mathcal{E} : j \in C\}| < k_i/2\}\right|}{|C|}$, is the fraction of nodes inside a neighborhood that have less than half of their edges pointing inside (internal+external quality, non-attributed).
- **OddBall** [1] uses a linear model to find neighborhoods that deviate in node density (internal consistency only, non-attributed).

- **SODA** [11] finds a max-margin hyperplane that separates connected and disconnected nodes using both structure and attributes. It ranks neighborhoods by the negative margin of this hyperplane (internal+external quality, attributed).
- **Attribute-Weighted Normalized Cut** is based on a cluster quality measure proposed by [24] for attributed graphs. They identify a subspace of attributes for a cluster, which minimizes its weighted normalized cut, where edges are weighted by the similarity of end-nodes on the selected subspace. Subspace selection is *quadratic* in the number of all attributes. Our real-world datasets DBLP, Citeseer, and LastFM have more than 23 thousand, 206 thousand, and 3.9 million attributes, respectively, for which [24] is intractable. As such, we consider a simplified version; by using a uniform weight vector over the full attribute space to compute normalized cut (internal+external quality, attributed).

## C Related Work

We organize related work into two groups: (1) analysis of community structure and community quality, and (2) anomaly detection in non-attributed and attributed graphs.

**C.1 Analysis of community structure and quality**
Leskovec *et al.* studied the statistical properties of communities in real social and information networks and analyzed how they split into communities and how typical community qualities (according to the conductance measure) change over a range of size scales [27]. Their findings suggest the absence of large well-defined communities, which has been corroborated in a later study by Gleich *et al.* [23]. Arnaboldi *et al.* analyzed the structure of ego networks and found that social relationships in online and offline social networks are organized similarly [19]. Akoglu *et al.* found that egonet characteristics display power-law-like patterns in real networks [18]. Other works studied the structure and dynamics of real-world graphs at large, without specific focus on communities [26, 31]. The focus of all these works is on networks with no attributes.

Several other works quantify the quality of communities in graphs. Yang and Leskovec investigated a long list of such measures and compared their performance based on ground-truth communities [43]. Newman studied the mixing properties in attributed graphs to quantify the correlations between the attributes of adjacent nodes; high correlation is referred as assortative mixing, which tends to break the network into communities [36]. Similarly, Silva *et al.* studied the correlation between attribute sets and dense subgraphs, called the structural correlation patterns in attributed graphs [42]. Finally, work in subspace clustering [34] has also been extended as a quality measure to find graph-cuts which correlate with attributes [24]. These methods have utilized their measures to define a global objective for the graph clustering/partitioning problem, and are not directly applicable to anomaly detection of neighborhoods.

**C.2 Graph anomaly detection** In their seminal work, Noble and Cook [39] used frequent subgraph mining and information theoretic principles to identify anomalous subgraphs in graphs with a *single* attribute. Similarly, Ni *et al.* developed algorithms to find sets of connected nodes in a graph for which a *single* attribute value is significantly higher than in the neighborhood of the set [28, 29]. Akoglu *et al.* proposed OddBall to find *structural* anomalies, where they found and used patterns in egonets to flag the anomalies in non-attributed graphs [18]. Gao *et al.* formulated a new problem to identify community outliers which deviate in the full attribute space from others that belong to the same community [22]. Perozzi *et al.*, on the other hand, focused on community outliers that deviate on a pre-defined subset of attributes that is inferred from user preference [40]. Both of those works find node outliers within communities rather than anomalous communities.

Most recently, Gunnemann *et al.* generalized the normalized cut measure for clustering attributed graphs [24]. Each cluster $k$ is associated with a respective subspace $s_k$ of relevant attributes, similar to our focus attributes. As such, a score called normalized subspace cut (NSC) can be computed for each cluster. While NSC score can be used as a quality measure for a given neighborhood, there are three main shortcomings to computing it: (1) The formulated objective is *not convex* in subspace $s_k$. (2) Proposed method is a *heuristic*, as the problem is not tractable for large graphs. Moreover, to make the search space more tractable, the authors limit solution vectors to binary: irrelevant/relevant, and by doing so compromise from inferring attribute weights. Finally, (3) despite the way the search space is limited to binary weight vectors, their proposed heuristic is *quadratic* in the number of all attributes $|S|$ (and not the subspace size $|s_k|$). This disables us to use their method for graphs with millions, or even tens of thousands of attributes (e.g., our LastFM and DBLP graphs in Table 1).

Most similar to ours is the work by Gupta *et al.* on outlier subgraph discovery [25]. Their formulation involves inferring an attribute subspace in which the margin between minimum dissimilarity among disconnected nodes and maximum dissimilarity among connected nodes is maximized. For normal subgraphs, this margin is expected to be large—as the minimum dissimilarity among disconnected nodes is still large and the maximum dissimilarity among connected nodes is small. Our quality formulation is considerably different and yields a much faster optimization. Moreover, none of the existing works including [25] has a notion of edge "exoneration" as we introduce in our work.

# References

[18] L. Akoglu, M. McGlohon, and C. Faloutsos. Oddball: Spotting anomalies in weighted graphs. In *PAKDD*, 2010.

[19] V. Arnaboldi, M. Conti, A. Passarella, and F. Pezzoni. Analysis of ego network structure in online social networks. In *SocialCom*, 2012.

[20] V. Blondel, J. Guillaume, R. Lambiotte, and E. Mech. Fast unfolding of communities in large networks. *Jour. Stat. Mech.*, 2008.

[21] A. Clauset, M. Newman, and C. Moore. Finding community structure in very large networks. *Physical Review E*, 70(6), 2004.

[22] J. Gao, F. Liang, W. Fan, C. Wang, Y. Sun, and J. Han. On community outliers and their efficient detection in information networks. In *KDD*, pages 813–822, 2010.

[23] D. F. Gleich and C. Seshadhri. Vertex neighborhoods, low conductance cuts, and good seeds for local community methods. In *KDD*, pages 597–605, 2012.

[24] S. Gunnemann, I. Farber, S. Raubach, and T. Seidl. Spectral subspace clustering for graphs with feature vectors. In *ICDM*, 2013.

[25] M. Gupta, A. Mallya, S. Roy, J. H. D. Cho, and J. Han. Local learning for mining outlier subgraphs from network datasets. In *SIAM SDM*, pages 73–81, 2014.

[26] J. Leskovec, J. Kleinberg, and C. Faloutsos. Graphs over time: densification laws, shrinking diameters and possible explanations. In *KDD*, pages 177–187. ACM Press, 2005.

[27] J. Leskovec, K. J. Lang, A. Dasgupta, and M. W. Mahoney. Statistical properties of community structure in large social and information networks. In *WWW*, pages 695–704, 2008.

[28] N. Li, Z. Guan, L. Ren, J. Wu, J. Han, and X. Yan. gIceberg: Towards iceberg analysis in large graphs. In *ICDE*, pages 1021–1032, 2013.

[29] N. Li, H. Sun, K. Chipman, J. George, and X. Yan. A probabilistic approach to uncovering attributed graph anomalies. In *SIAM SDM*, pages 82–90, 2014.

[30] D. S. Massey and N. A. Denton. The dimensions of residential segregation. *Social Forces*, 67(2):218–315, 1988.

[31] M. McGlohon, L. Akoglu, and C. Faloutsos. Weighted graphs and disconnected components: patterns and a generator. In *KDD*, pages 524–532, 2008.

[32] M. McPherson, L. Smith-Lovin, and J. M. Cook. Birds of a feather: Homophily in social networks. *Annual Review of Sociology*, 27(1):415–444, 2001.

[33] J. Moody. Race, school integration, and friendship segregation in america. *American J. of Sociology*, 107(3):679–716, 2001.

[34] E. Müller, S. Günnemann, I. Assent, and T. Seidl. Evaluating clustering in subspace projections of high dimensional data. *VLDB*, 2(1):1270–1281, 2009.

[35] M. Newman and M. Girvan. Mixing patterns and community structure in networks. In *Statistical Mechanics of Complex Networks*, volume 625, pages 66–87. 2003.

[36] M. E. J. Newman. Assortative mixing in networks. *Physical Review Letters*, 89(20), 2002.

[37] M. E. J. Newman. Modularity and community structure in networks. *PNAS*, 103(23):8577–8582, 2006.

[38] M. E. J. Newman. *Networks: An Introduction*. Oxford University Press, Oxford; New York, 2010.

[39] C. C. Noble and D. J. Cook. Graph-based anomaly detection. In *KDD*, pages 631–636. ACM, 2003.

[40] B. Perozzi, L. Akoglu, P. Iglesias Sánchez, and E. Müller. Focused clustering and outlier detection in large attributed graphs. In *KDD*, pages 1346–1355, 2014.

[41] H. Shiokawa, Y. Fujiwara, and M. Onizuka. Fast algorithm for modularity-based graph clustering. In *AAAI*, 2013.

[42] A. Silva, W. M. Jr., and M. J. Zaki. Mining attribute-structure correlated patterns in large attributed graphs. *PVLDB*, 5(5):466–477, 2012.

[43] J. Yang and J. Leskovec. Defining and evaluating network communities based on ground-truth. In *ICDM*, 2012.